\documentclass[aps,prm,superscriptaddress,twocolumn]{revtex4-1}
\usepackage[english]{babel}
\usepackage{times}
\usepackage{graphicx}
\usepackage{graphics}
\usepackage{amsmath}
\usepackage{mathtools}
\usepackage{amsfonts}
\usepackage{amssymb}
\usepackage{dsfont}
\usepackage{epstopdf}
\usepackage{makeidx}
\usepackage{subfigure}
\usepackage{color}
\usepackage{pgf}
\usepackage{bm}
\usepackage{ulem}
\usepackage{tikz}

\begin{document}

\title{Hedgehog orbital texture in $p-$type tellurium and the antisymmetric nonreciprocal Hall response}

\author{Gabriele Maruggi} 

\affiliation{Instituto de Fisica, Universidade Federal do Rio de Janeiro, Caixa Postal
68528, Rio de Janeiro, Brazil}

\author{Jaime Ferreira}

\affiliation{Centro Brasileiro de Pesquisas Físicas, Rua Dr. Xavier Sigaud, Rio de Janeiro, Brazil}

\author{Elisa Baggio-Saitovitch}

\affiliation{Centro Brasileiro de Pesquisas Físicas, Rua Dr. Xavier Sigaud, Rio de Janeiro, Brazil}

\author{Carsten Enderlein} 

\affiliation{Instituto de Fisica, Universidade Federal do Rio de Janeiro, Caixa Postal
68528, Rio de Janeiro, Brazil}

\author{Marcello B. Silva Neto} 

\affiliation{Instituto de Fisica, Universidade Federal do Rio de Janeiro, Caixa Postal
68528, Rio de Janeiro, Brazil}

\date{\today} 

\begin{abstract}

Tellurium is a gyrotropic, $p-$type Weyl semiconductor with remarkable electronic, optical, and transport properties. It has been argued that some of these properties might stem from Weyl nodes at crossing points in the band structure, and their nontrivial topological textures. However, Weyl nodes in time-reversal invariant semiconductors are split up in energy, rather than in momentum, and located deep below (far above) the top (bottom) of the valence (conduction) band, challenging such an interpretation. Here, instead, we use a $4-$band ${\bf k}\cdot{\bf p}$ Hamiltonian for $p-$type tellurium to show how the ${\bf k}-$dependent spin-orbit interaction mixes up the top two (Weyl node free) and bottom two (Weyl node containing) valence bands, generating a 3D hedgehog orbital magnetic texture at the uppermost valence band, accessible to transport already at the lowest doping. Hedgehog textures are important signatures of Weyl fermion physics in general and in the context of condensed matter physics arise form the carriers' wave packet rotation being locked to their propagation wavevector. For spatially dispersive media, such an induced hedgehog texture/carrier rotation stabilizes two novel, nonreciprocal and antisymmetric components to the Hall transport within different weak-localization (antilocalization) relaxation regimes: the anomalous and planar Hall effects, usually forbidden by time reversal symmetry. Our AC magnetotransport measurements on Sn-doped tellurium confirm the theoretical predictions and our work demonstrates how Weyl signatures generally appear in transport on enantiomorphic materials with natural optical activity.

\end{abstract}


\maketitle

\section{Introduction}

The discovery of topological insulators (TIs) \cite{TopologicalInsulators} paved the way for the emergence of a new class of remarkable quantum systems: topological materials (TMs) \cite{TopologicalMaterials,ReviewYoichi}. One of the earliest manifestations of topology in TMs is the integer quantum Hall effect (IQHE) \cite{IQHE,Halperin1982}, in which the Hall conductance plays the role of an adiabatic curvature whose associated topological invariant corresponds to the number of completely filled bands \cite{TKKN}. When conduction and valence bands linearly cross at the Fermi level one speaks of topological semimetals \cite{TopologicalSemiMetals}. Time-reversal symmetry breaking splits up the two bands in momentum introducing a chiral shift \cite{BurkovPRL2014}, thus generating two Weyl nodes that act as source and drain of nonzero Berry curvature. In such Weyl semimetals (WSMs) the chiral anomaly gives rise to open Fermi surfaces or Fermi arcs connecting the two Weyl nodes, leading to unique signatures in transport \cite{WeylSemiMetals} such as the anomalous Hall effect (AHE) which in WSMs is purely intrinsic and determined solely by the location of the linearly dispersing Weyl nodes \cite{Burkov2018}. Conveniently, the gauge/gravity duality provides a simple and elegant way to compute the AHE in bulk topological WSMs because of its holographic equivalence to a nonzero axial component for the horizon gauge field at the conformal boundary \cite{Gauge-Gravity-Holography}. The AHE is universal even for a doped Weyl metal when the Fermi level occurs slightly away from the nodes \cite{BurkovPRL2014}.   

Quite recently a novel class of TMs was discovered: Weyl semiconductors (WSCs) \cite{Zhang2020}. The unique signatures of Weyl physics found in the enantiomorphic, elemental semiconductor tellurium, include negative longitudinal magnetoresistence,  planar Hall effect (PHE), and logarithmically periodic magneto-oscillations \cite{Zhang2020}. Following the discovery, tellurium resurrected as a very promising material for both scientific research and the design of novel technologies \cite{ResurrectionTe}. Tellurium can be easily synthesized either as 1D nanowires, with promising piezoelectric and thermoelectric properties \cite{Te1D}, grown as 2D films, fulfilling all the necessary requirements to replace Silicon and Si-based semiconductor technology \cite{Te2D}, or as 3D bulk crystals, exhibiting quite remarkable chiral thermoelectric properties \cite{Te3D} and pressure induced metal insulator transition \cite{JaimeCommMat}. Tellurium also exhibits several magnetoelectric effects including current-induced magnetization \cite{CurrentInducedMagnetization}, strong electrical magneto-chiral anisotropy, circular photogalvanic effect, nonlinear (current induced) anomalous Hall effect, kinetic Faraday effect, kinetic magnetoelectric effect \cite{Sahin2018}, and natural optical activity \cite{Tsirkin2018}, that have all been interpreted as clear and unique signatures of Weyl physics. Finally, the chiral-induced spin selectivity (CISS) in tellurium \cite{Shalygin2012} builds an important bridge to systems that combine chiral molecules, such as amino-acids, sugars or DNA, and magnetic moments of charge carriers, therefore engineering new approaches to enantioselective chemistry \cite{EnantioSelectivityChemistry}.

Of particular relevance to understanding the origins of the Weyl physics in WSCs is the Hall response. For a magnetic field, ${\bf B}$, at an angle $\varphi$ with an applied current, ${\bf j}$, and both co-planar to the Hall electric field, ${\bf E}$, a quadratic, reciprocal PHE, $\sigma_{xy}^{ph}(B)\propto B^2\sin(2\varphi)$, is observed for $\varphi\neq 0,\pi/2$ in tellurium \cite{Zhang2020}. Although a symmetric PHE is ubiquitous in WSMs \cite{PHEWeylSemiMetals}, no signatures of an AHE \cite{BurkovPRL2014} has yet been reported. The Weyl physics in WSCs must then come from pairs of nodes split up in energy, not in momentum, that should also account for any antisymmetric, nonreciprocal, anomalous or planar Hall responses \cite{OnsagerNonReciprocal}, for ${\bf j}\perp{\bf B}$ ($\varphi=\pi/2$) and with ${\bf B}$ at an angle $\pi/2-\theta$ with ${\bf E}$. However, even though tellurium is always $p-$doped due to naturally occurring vacancies, such Weyl nodes are deep below the top of the valence band \cite{Zhang2020, Hirayama2015}, making any contribution to transport negligible. 

In this work we use a $4-$band ${\bf k}\cdot{\bf p}$ Hamiltonian \cite{Doi1970,Nakao1971} to demonstrate how the ${\bf k}-$dependent spin-orbit interaction (${\bf k}-$SOI) induces nonzero Berry curvatures at the top two, topologically trivial valence bands resulting from their mixing to the bottom two, topologically nontrivial and Weyl-node containing valence bands. The resulting form of wave-packet rotation undergone by the carriers \cite{XiaoRMP2010}  generates a 3D hedgehog orbital texture and contributes to transport. Using Boltzmann transport equations \cite{Kim2014,Johansson2019} for spatially dispersive media \cite{NagaosaNonreciprocal} and ${\bf j}\parallel\hat{\bf z}\perp{\bf B}\not\parallel{\bf E}\parallel\hat{\bf x}$, we have found, in addition to the reciprocal Hall response, $\sigma_{zx}^{h}(B)\propto B\cos\theta$, two novel, antisymmetric, nonreciprocal phenomena: the AHE, $\tilde{\sigma}_{zx}^{ah}(B)\propto B\cos\theta$ and the PHE $\tilde{\sigma}_{zx}^{ph}(B)\propto B\sin\theta$. Our AC magnetotransport for $p-$doped Te:Sn confirmed both the antisymmetry and phase relations amongst $\sigma^{h}_{zx},\tilde{\sigma}^{ah}_{zx}$, and $\tilde{\sigma}^{ph}_{zx}$.

\section{Tellurium}

\begin{figure}[htbp]
\includegraphics[scale=0.38]{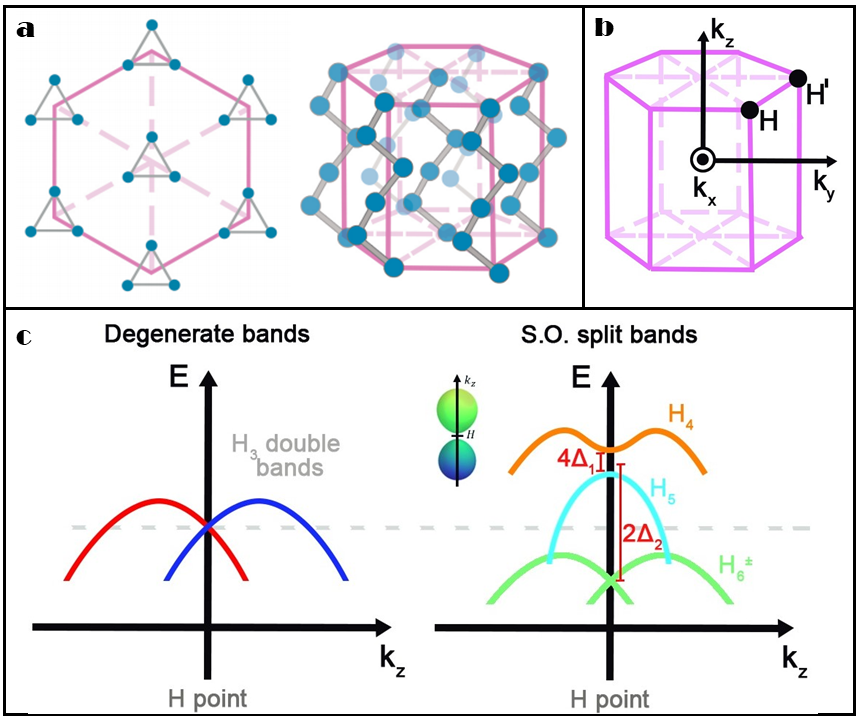}
\caption{$a -$ top and side views of the crystal structure of trigonal tellurium showing the arrangement of the helices; $b -$ the Brillouin zone and the $H$ and $H^\prime$ points; $c -$ valence band structure of tellurium and degeneracy lifting by the spin-orbit interaction.}
\label{FigureCrystalBandStructures}
\end{figure}

Elemental tellurium is a chiral, non-centrosymmetric semiconductor composed by helical chains arranged in a triangular lattice as shown in Fig. \ref{FigureCrystalBandStructures}. 
The space group of tellurium is either $D_3^4$ for right-handed or $D_3^6$ for left-handed screw axis \cite{Firsov1957}. When choosing as set of primitive translation vectors
$\mathbf{t}_1=(a,0,0)$, $\mathbf{t}_2=(-a/2,\sqrt{3}a/2,0)$, $\mathbf{t}_3=(0,0,c)$,
where $a=4.44\AA$ and $c=5.91\AA$, one can then write down the symmetry elements of the factor group in the space group of $D_3^6$.

\subsection{${\bf k}\cdot{\bf p}-$Hamiltonian}

The valence band structure of tellurium has been derived using the ${\bf k}\cdot{\bf p}$ perturbation theory \cite{Doi1970,Nakao1971}. One assumes the energy spectrum, $E_n({\bf k}_H)$, and wave-functions, $\psi_{n,{\bf k}_H}({\bf r})$, for the $n-$th valence band at ${\bf k}_H=(4\pi/3a,0,\pi/c)$, are solutions to the eigenvalue problem, ${\cal H}_0\psi_{n,{\bf k}_H}({\bf r})=E_n({\bf k}_H)\psi_{n,{\bf k}_H}({\bf r})$, for an unperturbed
${\cal H}_0=\mathbf{p}^2/2m+U(\mathbf{r})$, where $U(\mathbf{r})$ is the periodic potential. Then one looks 
for solutions to the Schrödinger's equation ${\cal H}\psi_{n,{\bf k}}({\bf r})=E_n({\bf k})\psi_{n,{\bf k}}({\bf r})$
for ${\bf k}$ away from the $H-$point where $E_n({\bf k})$ and $\psi_{n,{\bf k}}({\bf r})$ can be obtained from a double perturbation theory in ${\bf k}\cdot{\bf p}$ and the SOI. 
One inserts $\psi_{n,{\bf k}}({\bf r})=e^{i{\bf k}\cdot{\bf r}}u_{n,{\bf k}_H,{\bf k}}({\bf r})$ into Schrödinger's equation to obtain an eigenvalue problem for $u_{n,{\bf k}_H,{\bf k}}({\bf r})$ as
${\cal H}_{kp,so}u_{n,{\bf k}_H,{\bf k}}({\bf r})=E^\prime_n({\bf k}_H+{\bf k})u_{n,{\bf k}_H,{\bf k}}({\bf r})$, where $E^\prime_n({\bf k}_H+{\bf k})=E_n({\bf k}_H+{\bf k})-\hbar^2{\bf k}^2/2m$. 
The matrix elements of
${\cal H}={\cal H}_0+{\cal H}_1+{\cal H}_2+{\cal H}_3$, with 
\begin{eqnarray}
{\cal H}_1&=&\frac{\hbar}{m}{\bf k}\cdot{\bf p},\nonumber\\
{\cal H}_2&=&\frac{\hbar}{4m^2c^2}(\mathbf{\sigma}\times\nabla U(\mathbf{r}))\cdot\mathbf{p},\nonumber\\ 
{\cal H}_3&=&\frac{\hbar^2}{4m^2c^2}(\mathbf{\sigma}\times\nabla U(\mathbf{r}))\cdot\mathbf{k},
\end{eqnarray}
are calculated in terms of the unperturbed states $\psi_{n,{\bf k}_H}({\bf r})$, around ${\bf k}={\bf k}_H$, and correspond, respectively, to contributions from ${\bf k}\cdot{\bf p}$, from the ${\bf k}-$independent SOI, and from the ${\bf k}-$SOI \cite{Doi1970,Nakao1971}.
%
%
The doubly degenerate $H_3$ valence band state becomes split into $H_6^\pm,H_4^+$, and $H_5^-$ valence bands by the spin-orbit interaction \cite{Doi1970,Nakao1971}. 
Neglecting ${\cal H}_3$ our starting-point Hamiltonian reads \cite{Doi1970,Nakao1971}
\begin{equation}
\begin{bmatrix}
\varepsilon_{++}({\bf k})-\Delta_2 & a k_+ & -iRk_- & ib^*k_- \\
a k_- & \varepsilon_{-+}({\bf k})-\Delta_2 & -ib^*k_+ & iRk_+  \\
iRk_+ & ibk_- & \varepsilon_{--}({\bf k})+\Delta_2 & -2\Delta_1 \\
-ibk_+ & -iRk_- & -2\Delta_1 & \varepsilon_{+-}({\bf k})+\Delta_2
\end{bmatrix}.\nonumber
\end{equation}
The $k-$independent parameters $\Delta_{1,2}$ arise at first-order in ${\cal H}_2$ and are responsible for the bands splitting. At second-order in ${\cal H}_1$ we have $\varepsilon_{\pm,\pm}({\bf k})=\varepsilon({\bf k}) \pm(S\pm d)k_z$, with $\varepsilon({\bf k})=A k_\perp^2+B k_z^2$ \cite{Doi1970,Nakao1971}. The $S k_z$ and $R k_\pm$ terms represent linear-in-$k$ contributions at first-order in ${\cal H}_1$ due to the lack of inversion symmetry. Combining ${\cal H}_1$ and ${\cal H}_2$ gives rise to the linear-in-$k$ band mixing elements $a k_\pm$ and $i b k_\pm$, as well as the diagonal $\pm d k_z$ terms. We have discarded all trigonal warping terms $k_\pm^2$ and $(k_x^2-k_y^2)$, and all other second-order terms in ${\cal H}_2$ \cite{Doi1970,Nakao1971}. 

\subsection{Valence band structure}

The band structure obtained from the above Hamiltonian, for $R=b=0$, at the $H-$point is shown in Fig. \ref{FigureCrystalBandStructures}$-c$ on the right, where 
\begin{eqnarray}
    \varepsilon_{H_4^+,H_5^-}({\bf k})&=&\varepsilon({\bf k})\pm\sqrt{(S-d)^2k_z^2+4\Delta_1^2}+\Delta_2, \nonumber \\
    \varepsilon_{H_6^\pm}({\bf k})&=&\varepsilon({\bf k})\pm\sqrt{(S+d)^2k_z^2+a^2k_\perp^2}-\Delta_2.
\end{eqnarray}
The uppermost $H_4^+$ valence band exhibits a dumbbell-shaped structure at
$k_0^\pm=\pm\sqrt{(S-d)^4-16\Delta_1^2B^2}/2B|S-d|$, as long as $|(S-d)^2/B|>4\Delta_1$, which is the case for tellurium. The normalized eigenstates for $R=b=0$ are
\begin{eqnarray}
\left|u_{H_{4,5}^{\pm}}({\bf k})\right>&=&\frac{1}{\sqrt{2 h({\bf k})(h({\bf k})\pm h_z(k_z))}}
\left(
\begin{matrix}
2\Delta_1 \\ 
-h_z(k_z) \mp h({\bf k})
\end{matrix}
\right),\nonumber\\
\left|u_{H_{6}^{\pm}}({\bf k})\right>&=&\frac{1}{\sqrt{2 \mathfrak{g}({\bf k})(\mathfrak{g}({\bf k})\mp\mathfrak{g}_z(k_z))}}
\left(
\begin{matrix}
-a k_+ \\ 
\mathfrak{g}_z(k_z)\mp \mathfrak{g}({\bf k})\nonumber
\end{matrix}
\right),
\label{Eigenstates}
\end{eqnarray}
with $h_z(k_z)=(S-d)k_z$ and $\mathfrak{g}_z(k_z)=(S+d)k_z$, while $h({\bf k})=\sqrt{h_z^2(k_z)+4\Delta_1^2}$ and $\mathfrak{g}({\bf k})=\sqrt{\mathfrak{g}_z^2(k_z)+ak_\perp^2}$. 

\section{Topological Properties}

\subsection{Berry curvature $-$ $\mathbf{\Omega}({\bf k})$}

In this section we will demonstrate how a nonzero Berry curvature can be induced at a Weyl node free valence band in second order perturbation theory due to its coupling to Weyl node containing valence bands. The Berry curvature of the $n-$th band can be calculated from the expression \cite{XiaoRMP2010}
\begin{equation}
    \Omega_n^\alpha({\bf k})=i\sum_{n\neq n^\prime}\sum_{\mu\nu}\epsilon^{\alpha\mu\nu}
    \frac{\langle u_n\left|\partial_\mu {\cal H}\right| u_{n^\prime}\rangle
    \langle u_{n^\prime}\left|\partial_\nu {\cal H}\right| u_n\rangle}{(\varepsilon_n-\varepsilon_{n^\prime})^2},
    \label{GeneralBerry}
\end{equation}
where $n,n^\prime=H_4^+,H_5^-,H_6^\pm$, derivaties are with respect to $\partial_\mu\equiv\partial/\partial k^\mu$, and $\epsilon^{\alpha\mu\nu}$ is the total anti-symmetric tensor with $\alpha,\mu,\nu=x,y,z$. Using $R=b=0$ we find  $\mathbf{\Omega}_{H_4^+,H_5^-}({\bf k})=(0,0,0)$, showing that, at the $H$ or $H^\prime$ points, the $H_4^+$ and $H_5^-$ bands are topologically trivial and carry zero Berry curvature. In contrast, the same calculation for the lowest two bands yields
$\mathbf{\Omega}_{H_6^\pm}({\bf k})=(0,0,\Omega_{H_6^\pm}^z({\bf k}))$, 
with
\begin{equation}
    \Omega_{H_6^\pm}^z({\bf k})=\pm\frac{1}{2}a^2\frac{\mathfrak{g}_z(k_z)}{\mathfrak{g}^3({\bf k})},
\end{equation}
which is simply the field of a magnetic monopole at the origin in momentum space \cite{XiaoRMP2010}. This shows that, at the $H$ or $H^\prime$ points, the $H_6^\pm$ valence bands are indeed topological, a direct consequence of the Weyl node at ${\bf k}_H$ \cite{Hirayama2015}. 

For deviations from the $H$ point, however, the bands mixing may no longer be ignored. By considering $b\neq 0$ and $R\neq 0$ the induced curvature at the $H_4^+$ band due to the mixing with the $H_6^\pm$ bands reads
$\mathbf{\Omega}^{(0)}_{H_4^+}({\bf k})=(0,0,\Omega_{H_4^+}^{(z,0)}({\bf k}))$,
where
\begin{widetext}
\begin{equation}
     \Omega_{H_4^+}^{(z,0)}({\bf k})=\sum_{\sigma=\pm}
     \frac{R^2+|b|^2}{(h({\bf k})-\sigma \mathfrak{g}({\bf k})+2\Delta_2)^2}
     \frac{[h_z(k_z)+h({\bf k})]^2[\mathfrak{g}^2({\bf k})-\mathfrak{g}^2_z(k_z)]-
     [h^2({\bf k})-h_z^2(k_z)][\mathfrak{g}_z(k_z)-\sigma\mathfrak{g}({\bf k})]^2}
     {2\mathfrak{g}({\bf k})h({\bf k})(h({\bf k})+h_z(k_z))(\mathfrak{g}({\bf k})-\sigma\mathfrak{g}_z(k_z))}.
\end{equation}
\end{widetext}
The above result demonstrates that, by considering solely the ${\bf k}-$independent SOI to the $H_{6,6^\prime}$ bands a $z-$component for the Berry curvature can be induced at the $H_4^+$ valence band. Here the superscript $(z,0)$ indicates that the $z-$component was calculated {\it without} considering the ${\bf k}-$SOI.

However, so far, we have omitted all matrix elements associated to the ${\bf k}-$SOI in ${\cal H}_3$ \cite{Doi1970,Nakao1971} and these are the central new theoretical ingredient of the present work, since these will induce nonzero contributions to the other two components of the Berry curvature. Combining ${\cal H}_1$ and ${\cal H}_3$ gives rise to the following perturbation Hamiltonian
\begin{equation}
\small
\begin{bmatrix}
0 & 0 & (G-i u) k_-k_z & -v k_zk_-\\
0 & 0 & -vk_zk_+ & (G+i u) k_+k_z \\
(G+i u) k_+k_z & -v^*k_zk_- & 0 & 0 \\
-v^*k_zk_+ & (G-i u) k_-k_z & 0 & 0
\end{bmatrix}.\nonumber
\end{equation}
We use again the general expression for the Berry curvature (\ref{GeneralBerry}) to calculate the fully induced curvature onto the $H_4^+$ band 
\begin{equation}
    \mathbf{\Omega}^{(1)}_{H_4^+}({\bf k})=(\Omega_{H_4^+}^{(x,1)}({\bf k}),\Omega_{H_4^+}^{(y,1)}({\bf k}),\Omega_{H_4^+}^{(z,1)}({\bf k})),
\end{equation}
where now 
\begin{widetext}
\begin{eqnarray}
     \Omega_{H_4^+}^{(x,1)}({\bf k})&=&\sum_{\sigma=\pm}
     \frac{(G^2+|u|^2) k_x k_z}{(h({\bf k})-\sigma \mathfrak{g}({\bf k})-2\Delta_2)^2}
     \frac{[h_z({\bf k})+h({\bf k})]^2[\mathfrak{g}_z({\bf k})-\sigma\mathfrak{g}({\bf k})]^2-
     [h^2({\bf k})-h_z^2(k_z)][\mathfrak{g}^2({\bf k})-\mathfrak{g}_z^2({\bf k})]}
     {2\mathfrak{g}({\bf k})h({\bf k})(h({\bf k})+h_z(k_z))(\mathfrak{g}({\bf k})+\sigma\mathfrak{g}_z(k_z))},\nonumber\\
     \Omega_{H_4^+}^{(y,1)}({\bf k})&=&\sum_{\sigma=\pm}
     \frac{(G^2+|u|^2) k_y k_z}{(h({\bf k})-\sigma \mathfrak{g}({\bf k})-2\Delta_2)^2}
     \frac{[h_z({\bf k})+h({\bf k})]^2[\mathfrak{g}_z({\bf k})-\sigma\mathfrak{g}({\bf k})]^2-
     [h^2({\bf k})-h_z^2(k_z)][\mathfrak{g}^2({\bf k})-\mathfrak{g}_z^2({\bf k})]}
     {2\mathfrak{g}({\bf k})h({\bf k})(h({\bf k})+h_z(k_z))(\mathfrak{g}({\bf k})+\sigma\mathfrak{g}_z(k_z))},\nonumber\\
     \Omega_{H_4^+}^{(z,1)}({\bf k})&=&\sum_{\sigma=\pm}
     \frac{(G^2+|u|^2) k_z^2}{(h({\bf k})-\sigma \mathfrak{g}({\bf k})-2\Delta_2)^2}
     \frac{[h_z({\bf k})+h({\bf k})]^2[\mathfrak{g}_z({\bf k})-\sigma\mathfrak{g}({\bf k})]^2-
     [h^2({\bf k})-h_z^2(k_z)][\mathfrak{g}^2({\bf k})-\mathfrak{g}_z^2({\bf k})]}
     {2\mathfrak{g}({\bf k})h({\bf k})(h({\bf k})+h_z(k_z))(\mathfrak{g}({\bf k})+\sigma\mathfrak{g}_z(k_z))}.
     \label{3DBerryCurvature}
\end{eqnarray}
\end{widetext}
The above result finally demonstrates that, by including also the ${\bf k}-$dependent SOI to the $H_{6,6^\prime}$ bands a full $3D$, radial Berry curvature is induced at the $H_4^+$ valence band. Here the superscript $(i,1)$ for all $i=x,y,z$ indicates that those components were calculated {\it including} the ${\bf k}-$SOI.

\subsection{Orbital texture $-$ ${\bf m}_{orb}({\bf k})$}

Knowledge of the original eigenstates of the $4-$band ${\bf k}\cdot{\bf p}-$Hamiltonian, obtained in eq. (\ref{Eigenstates}) allows us to calculate the induced orbital moment at the Weyl node free $H_4^+$ valence band due to its mixing to the two Weyl node containing $H_{6,6^\prime}$ valence bands. For that we shall use \cite{ChangNiu}
\begin{equation}
    m_{orb}^\alpha({\bf k})=\frac{ie}{2\hbar}\sum_{n\neq n^\prime}\sum_{\mu\nu}\epsilon^{\alpha\mu\nu}
    \frac{\langle u_n\left|\partial_\mu {\cal H}\right| u_{n^\prime}\rangle
    \langle u_{n^\prime}\left|\partial_\nu {\cal H}\right| u_n\rangle}{\varepsilon_n-\varepsilon_{n^\prime}},
    \label{OrbitalMoment}
\end{equation}
from which we can calculate the several contributions to the three components of the induced orbital texture. In order to simplify our analysis, and without loss of generality, we shall neglect the spin-splitting of the bands $H_6^\pm$ and the dispersion of the $H_4^+$ band, when compared to $\Delta_2$. Now if we consider both terms $b\neq 0$ and $R\neq 0$, we find a collinear orbital magnetic moment, ${\bf m}^{(0)}_{orb}=(0,0,m_{orb}^{(z,0)})$, with
\begin{equation}
    m_{orb}^{(z,0)}({\bf k})\approx
     \frac{e}{\hbar}\frac{R^2+|b|^2}{2\Delta_2}
     \frac{(S-d)k_z}{\sqrt{(S-d)^2 k_z^2+4\Delta_1^2}}.
     \label{z-orbital-moment}
\end{equation}
The existence of a nonzero $m_{orb}^{(z,0)}({\bf k})\sim k_z$ for the $H_4^+$ band in tellurium has been previously obtained and accounts for both the kinetic magnetoelectric effect observed in trigonal tellurium \cite{Sahin2018} and the current induced spin polarization of holes in tellurium \cite{Shalygin2012}. However, since at this stage $m_{orb}^{(x,0)}({\bf k})=m_{orb}^{(y,0)}({\bf k})=0$ it does not correspond to a Weyl fermion and we shall proceed with our analysis.

\begin{figure}[htbp]
\includegraphics[scale=0.32]{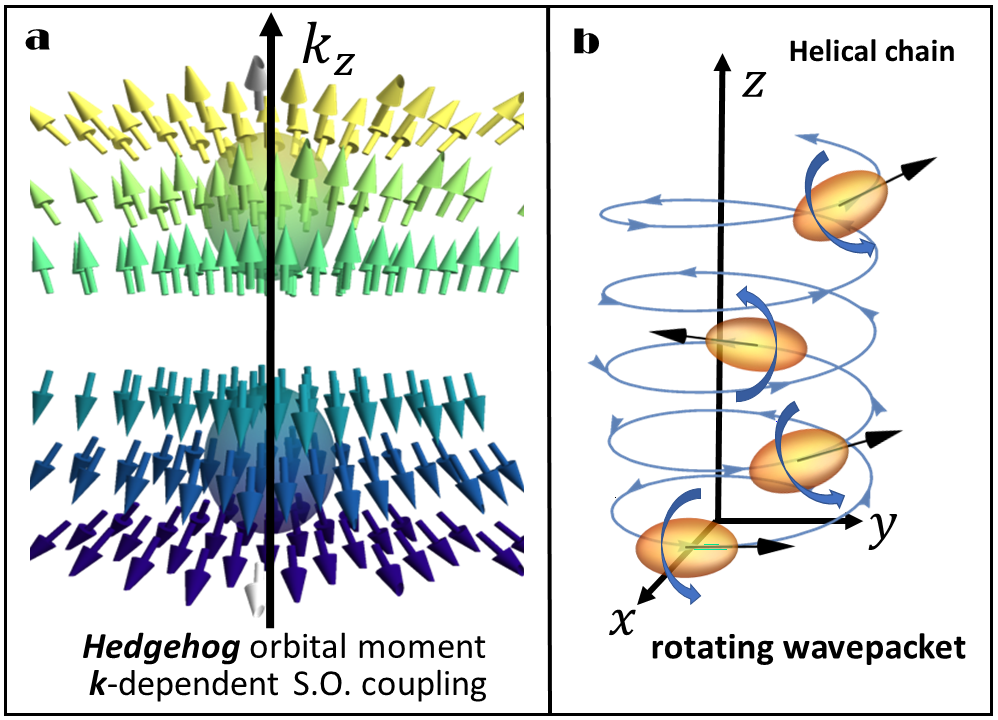}
\caption{Left $-$ Dumbbel Fermi surfaces and the hedgehog texture. For $k_z>0$ the Berry curvature is positive (yellow) while for $k_z<0$ the Berry curvature is negative (blue). Right $-$ Helical motion and wave packet rotation responsible for the hedgehog orbital texture.}
\label{FigureBerryCurvatureZ}
\end{figure}

Following the same reasoning used while discussing the Berry curvature, at this point we need to include all matrix elements associated to the ${\bf k}-$SOI in ${\cal H}_3$ \cite{Doi1970,Nakao1971}. The contributions from ${\cal H}_1$ and ${\cal H}_3$ gives rise to a full $3D$ orbital magnetic moment, ${\bf m}^{(1)}_{orb}({\bf k})=(m_{orb}^{(x,1)}({\bf k}),m_{orb}^{(y,1)}({\bf k}),m_{orb}^{(z,1)}({\bf k}))$, whose components are 
\begin{eqnarray}
     m_{orb}^{(x,1)}({\bf k})&\approx&
     \frac{e}{\hbar}\frac{G^2+|u|^2+|v|^2}{2\Delta_2}
     \frac{(S-d)k_z^2 k_x}{\sqrt{(S-d)^2 k_z^2+4\Delta_1^2}},\nonumber\\
     m_{orb}^{(y,1)}({\bf k})&\approx&
     \frac{e}{\hbar}\frac{G^2+|u|^2+|v|^2}{2\Delta_2}
     \frac{(S-d)k_z^2 k_y}{\sqrt{(S-d)^2 k_z^2+4\Delta_1^2}},\nonumber\\
     m_{orb}^{(z,1)}({\bf k})&\approx&
     \frac{e}{\hbar}\frac{G^2+|u|^2+|v|^2}{2\Delta_2}
     \frac{(S-d)k_z^2 k_z}{\sqrt{(S-d)^2 k_z^2+4\Delta_1^2}}.
     \label{3DOrbitalMoment}
\end{eqnarray}
It is important to emphasize that the 3D hedgehog orbital texture shown in Fig. \ref{FigureBerryCurvatureZ} {\it is not} to be associated to sources or sinks of Berry curvatures at $k_z=\pm k_0$, which are the positions of the two maxima of the uppermost $H_4^+$ valence band. This would mean Weyl nodes split in momentum, causing time-reversal symmetry to break down and it would be equivalent to the case of Weyl nodes in Weyl semimetals. Here, instead, the Weyl nodes are split up in energy, not in momentum, and the 3D hedgehog texture simply reflects the existence of Weyl nodes within the valence and conduction bands \cite{Hirayama2015} {\it at the same momentum} ${\bf k}_H$ or ${\bf k}_{H^\prime}$, and are therefore sources and sinks of Berry curvature at the same high-symmetry axis, either $H-$ or $H^\prime-$, in the Brillouin zone.

\section{Weyl Fermion Physics}

\begin{figure}[htbp]
\includegraphics[scale=0.3]{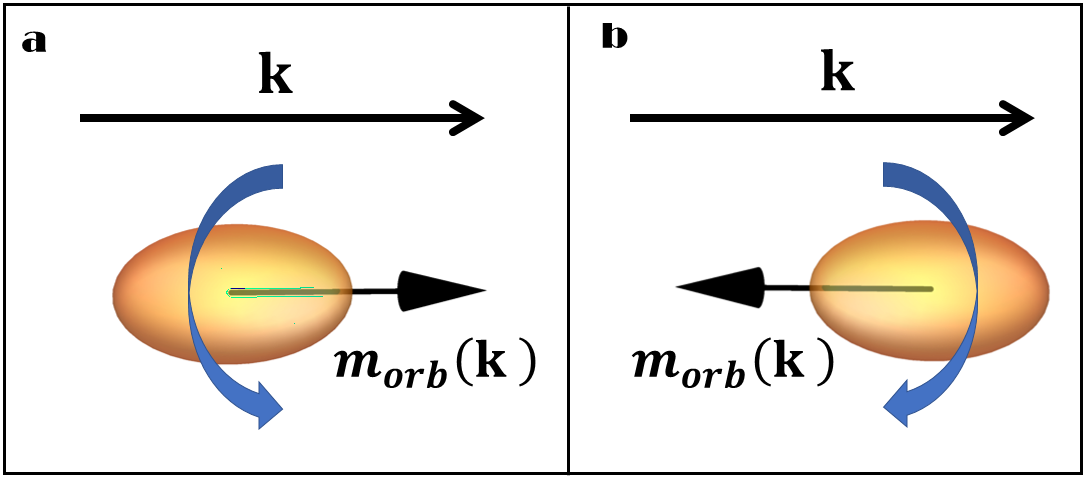}
\caption{$a -$ a right-handed Weyl fermion whose orbital angular momentum is parallel to its wavevector, ${\bf m}^+_{orb}({\bf k})\parallel +{\bf k}$; $b -$ a left-handed Weyl fermion whose orbital angular momentum is antiparallel to its wavevector, ${\bf m}^-_{orb}({\bf k})\parallel -{\bf k}$.}
\label{FigureWeylFermions}
\end{figure}

In particle physics, a Weyl fermion is a massless, spin $1/2$ particle whose spin is locked to its momentum, ${\bf s}^\pm({\bf p})\parallel\pm\vec{\bf p}$, defining its chirality: right-handed ($+$) or left-handed ($-$). This can be easily seen from  the Weyl Hamiltonian
\begin{equation}
H_{Weyl}=\pm\sigma\cdot{\bf p}.
\end{equation}
The solution to this Hamiltonian is straightforward. The eigenvalues define the linearly dispersing Weyl cone, $E_\pm({\bf p})=\pm|{\bf p}|$, and the eigenstates are
\begin{equation}
\left|u_{\pm}({\bf p})\right>=\frac{1}{\sqrt{2 |{\bf p}|(|{\bf p}|\mp p_z)}}
\left(
\begin{matrix}
\pm p_\perp \\ 
 (|{\bf p}|\mp p_z)e^{i\phi}
\end{matrix}
\right),
\end{equation}
where as usual $p_\perp=\sqrt{p_x^2+p_y^2}$ and $\phi=\arctan{(p_y/p_x)}$. The $z-$component of the Berry curvature associated to this Hamiltonian reads
\begin{equation}
b^z({\bf p})=\pm\frac{1}{2}\frac{p_z}{|{\bf p}|^3},
\end{equation}
representing a magnetic monopole at the origin, ${\bf p}=(0,0,0)$. The spin of the Weyl fermions can also be calculated
from ${\bf s}^\pm({\bf p})\equiv\langle u_{\pm}({\bf p})|\sigma|u_{\pm}({\bf p})\rangle$ and reads
\begin{equation}
{\bf s}^\pm({\bf p})=\pm\frac{1}{2}\frac{{\bf p}}{|{\bf p}|}.
\label{WeylFermion}
\end{equation}
This result shows us that indeed ${\bf s}^\pm({\bf p})$ is locked to the direction of the momentum $\pm{\bf p}$ with chirality right-handed ($+$) or left-handed ($-$) depending on the sign, see Fig. \ref{FigureWeylFermions}.

%
%

In condensed matter physics, charge carriers in, for example, $p-$type doped semiconductors are characterized by their nonzero effective masses, $m_{\perp,\parallel}^*\neq 0$, and parabollic dispersing valence bands. Nevertheless, Weyl fermion physics may arise from the interaction between these carriers to quasiparticle excitations close to Weyl-cone-like, linearly dispersing crossings in the band structure, in which case an intrinsic orbital angular momentum, associated to a wave packet rotation, might become locked to a propagation wavevector, ${\bf m}^\pm_{orb}({\bf k})\parallel\pm{\bf k}$, see Fig. \ref{FigureWeylFermions}. Comparing eqs. (\ref{3DOrbitalMoment}) and (\ref{WeylFermion}) we conclude that, indeed, the emergence of Weyl fermion physics in tellurium does occur and is a direct consequence of the ${\bf k}-$SOI between Weyl node free and Weyl node containing valence bands. The contribution from the ${\bf k}-$SOI to the orbital momentum in eq. (\ref{3DOrbitalMoment}) can be written as
\begin{equation}
{\bf m}^{\pm}_{orb}({\bf k})=\pm\frac{e}{\hbar}\left[\frac{G^2+|u|^2+|v|^2}{2\Delta_2}\frac{(S-d)k_z^2}{\sqrt{(S-d)^2 k_z^2+4\Delta_1^2}}\right]{\bf k},
\end{equation}
which has precisely the wavevector dependence expected for a Weyl fermion, as discussed in eq. (\ref{WeylFermion}). Here, the $\pm$ signs are associated to the two possible space groups of tellurium, either $D_3^4$ for right-handed $(+)$ or $D_3^6$ for left-handed $(-)$ screw axis \cite{Firsov1957}, and the $k_z$ dependence in the amplitude of ${\bf m}^{\pm}_{orb}({\bf k})$ is a consequence of the ${\bf k}-$dependent spin-orbit interaction.

\section{Onsager's relations in linear response}

\begin{figure}[htbp]
\includegraphics[scale=0.26]{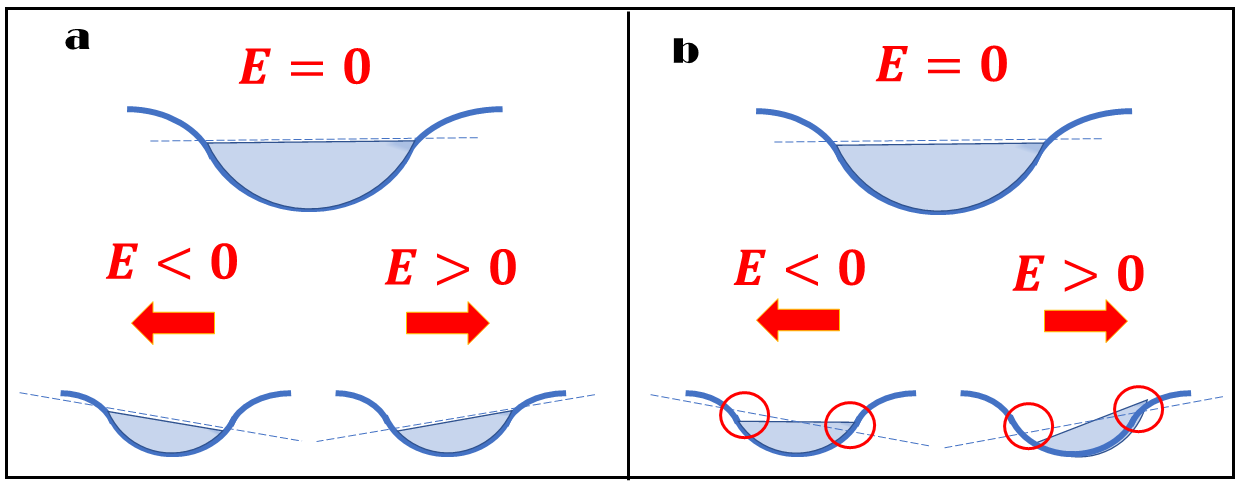}
\caption{{\bf a} $-$ Reciprocal response: the electric field varies slowly and the system follows adiabatically the perturbation via several relaxation processes such as impurity or lattice scattering. {\bf b} $-$ Nonreciprocal response: the electric field varies rapidly and the system fails to follow adiabatically the perturbation, independent of the various relaxation processes available. Nonreciprocal responses are very common in dissipative or strongly correlated systems, as well as in the case of gyrotropic media with natural optical activity, and lead to microscopic charge accumulation and spatial dispersion \cite{MorimotoNagaosa2018}.}
\label{FigureReciprocity}
\end{figure}

Ordinary matter, when unperturbed, flows towards an equilibrium state \cite{Hertel}. Such equilibrium state depends on the temperature, pressure, as well as several other external parameters such as: regions of space, containing a certain concentration of particles, mechanical stress, intensity of applied electric and magnetic fields, among others. When such parameters vary slowly, the system can return to its equilibrium state almost instantaneously and such process is said to be reversible, see for example Fig. \ref{FigureReciprocity}$-$a. When, instead, the variation is so fast that the system fails to adapt, it will remain out of equilibrium and the process is said to be irreversible. For example, when an external AC electric field varies so rapidly that the charge carriers contained inside micrometric regions in the sample fail to relax instantaneously, through some relaxation process such as scattering by impurities or by the lattice, then retardation will occur, see for example Fig. \ref{FigureReciprocity}$-$b. This is most commonly achieved in the presence of dissipation or strong correlations \cite{MorimotoNagaosa2018}. Retardation can be calculated in linear approximation and several relations for generalized susceptibilities can be derived, irrespective of the specific type of Hamiltonian, such as Kramers-Kronig relations, fluctuation-dissipation theorem, second law of thermodynamics and Onsager's relations. These relations in turn allow us to describe several transport, electro-optical, and magneto-optical effects: including Pockels effect, Faraday effect, Kerr effect, chiral magnetic effect, anomalous Hall effect \cite{Hertel}, and if we include also spatial dispersion, these relations can be readly generalized to include also natural optical activity, gyrotropic birrefringence, and several other non reciprocal phenomena.

The response of a system in equilibrium to external perturbations can be expressed in terms of generalized susceptibilities,
$\sigma_{ij}(\omega)$, which are matrices connecting the response $i$ to the perturbation $j$ \cite{Hertel}. In the case of adiabatic processes, when the external parameters vary very slowly, the system will return to equilibrium. In these static, $\omega=0$, cases we have  
\begin{equation}
    \sigma_{ij}(0)=\sigma_{ji}(0).
\end{equation}
This is Onsager's original relation for the case of generalized, static susceptibilities, as the one depicted in Fig. \ref{FigureReciprocity}$-$a. In order to extend such relation to the case of fast processes, $\omega\neq 0$, one needs the concept of time reversal \cite{Hertel}.

\subsection{Reciprocity and space homogeneity}

Let us discuss the behavior of the generalized susceptibilities, $\sigma_{ij}(\omega)$, for the electric field
\begin{equation}
    {\bf E}(\omega)={\bf E}_0\; e^{-i\omega t},
\end{equation}
describing a homogeneous perturbation in space. This approximation is valid when atomic dimensions, $a$, are negligible when compared to the wavelength, $\lambda$, of the electric field, $a\ll\lambda$. The requirement that the Hamiltonian should be even under time reversal \cite{Hertel} implies that, for fast processes and in the presence of an external magnetic field, one should have
\begin{equation}
    \sigma_{ij}(\omega,{\bf B})=\sigma_{ji}(\omega, -{\bf B}).
\end{equation}
We see that the symmetries restrict the components of the generalized susceptibility matrices and that, in the Ohmic regime of an isotropic, homogeneous system, they can be written as an expansion in powers of the applied magnetic field
\begin{equation}
    \sigma_{ij}(\omega,{\bf B})=\sigma(\omega)\delta_{ij}+h(\omega)\epsilon_{ijk} B_k+\dots,
    \label{ReciprocalOnsager}
\end{equation}
where $\delta_{ij}$ Kronecker's delta and $\epsilon_{ijk}$ is totally anti-symmetric Levi-Civita symbol. The first term describes the electrical conductivity inside the material in the absence of external magnetic fields. The second term needs to be anti-symmetric in $i,j$ due to the linear coupling to $B$, so that the Hamiltonian remains even under time reversal symmetry. The resulting electric current is thus necessarily perpendicular to both the applied magnetic field and to the applied external current, and this is the geometry that defines the conventional Hall effect. 

\subsection{Nonreciprocity and spatial dispersion}

Up to this point Onsager's relations have been discussed in the regime where atomic dimensions are negligible when compared to the wavelength of the electric field, $a\ll\lambda$, with
\begin{equation}
    {\bf E}({\bf q},\omega)={\bf E}_0\; e^{i({\bf q}\cdot{\bf x}-\omega t)}.
\end{equation}
The homogeneous limit, ${\bf q}=0$, corresponds to the zero-th order in a $a/\lambda$ expansion, with $\lambda=2\pi/|{\bf q}|$ \cite{LandauLifshitz}. However, in the macroscopic theory of electrodynamics the power expansion in $a/\lambda$ corresponds to an expansion in the displacement vector, ${\bf D}$, and not only in the electric field, ${\bf E}$, but also in its derivatives \cite{LandauLifshitz}
\begin{equation}
    D_i=\epsilon_{ik}E_k+\gamma_{ik\ell}\frac{\partial E_k}{\partial x_\ell},
\end{equation}
where $\epsilon_{ik}$ and $\gamma_{ik\ell}$ are functions of the frequency, $\omega$, and these tensors follow the symmetry principles of the kinetic coefficients in such a way that \cite{LandauLifshitz}
\begin{equation}
    \epsilon_{ik}=\epsilon_{ki},\quad\quad\gamma_{ik\ell}=-\gamma_{ki\ell}.
\end{equation}
This expansion can be extended to the conductivity tensor and should satisfy the extended Onsager's relations \cite{Hertel}
\begin{equation}
    \sigma_{ij}({\bf q},\omega,{\bf B})=\sigma_{ji}(-{\bf q},\omega, -{\bf B}).
\end{equation}
Once again power expansion of the external magnetic field and electromagnetic wavevector gives us
\begin{eqnarray}
    \sigma_{ij}({\bf q},\omega,{\bf B})&=&\sigma^{D}({\bf q},\omega)\delta_{ij}+\chi_{ijk}(\omega)q_k+
    h(\omega)\epsilon_{ijk}B_k\nonumber\\
    &+&\underbrace{g_{ijk\ell}(\omega)q_k B_\ell}_\text{nonreciprocal gyrotropy}+\dots,
    \label{NonRecOnsager}
\end{eqnarray}
where $\sigma^{D}({\bf q},\omega)$ is Drude's conductivity for an isotropic system and the tensors $\chi_{ijk}$ and $\epsilon_{ijk}$ are anti-symmetric in $i,j$. We have seen before that the term $h(\omega)\epsilon_{ijk}B_k$ describes the conventional Hall effect while the new term $\chi_{ijk}(\omega)q_k$ describes the phenomenon of natural optical activity. 

Most remarkable is the nonreciprocal gyrotropy contribution appearing in the second line of eq. (\ref{NonRecOnsager}). Here the tensor $g_{ijk\ell}$ is symmetric in the first two indices, $i,j$, in such a way that the new contribution $g_{ijk\ell}(\omega)q_k B_\ell$ allows for a linear term in the magnetic field contributing to the conductivity {\it without the necessity of being simultaneously perpendicular to the applied current and applied magnetic field}. This contribution was already known to  give rise to the phenomenon of the gyrotropic birefringence, the nonreciprocal propagation of waves inside the material. As we shall soon see, this is precisely the ${\bf q}-$dependent term that also makes room for a nonreciprocal linear planar Hall effect in tellurium. 

\section{Transport Properties}

\subsection{General inhomogeneous Weyl systems}

As we have learned from the previous discussion, nonreciprocity allows for the observation of a range of phenomena that would be otherwise forbidden in the reciprocal regime. Two of such phenomena are precisely the anomalous and the linear, antisymmetric planar Hall effects. Both are forbidden by time reversal symmetry in the reciprocal response, but may arise in the nonreciprocal response for spatially dispersive or inhomogeneous media. In order to see how this generally comes about, let us revisit the problem of Weyl fermions in the presence of inhomogeneities. This is usually done by minimally coupling Weyl fermions to spatially dispersive background gauge fields \cite{ChiralFermionsEWiMatter}, $A({\bf r})$ for the electromagnetic and ${\cal A}({\bf r})$ for the chiral, and whose ${\bf r}$ dependence is characterized by a modulation wavevector, ${\bf q}\neq 0$ \cite{AHEiQCD}. By moving from the laboratory to a local reference frame, through a local gauge transformation, one ends up with a Weyl$_{\bf q}$ Hamiltonian \cite{AHEiQCD}
\begin{equation}
H_{Weyl_{\bf q}}=
\begin{bmatrix}
(\tau_z/2)\sigma\cdot{\bf q} & \sigma\cdot{\bf p} \\
\sigma\cdot{\bf p} & (\tau_z/2)\sigma\cdot{\bf q}
\end{bmatrix},
\end{equation}
where $\tau_z$ is a pseudospin Pauli matrix. Once again the spectrum and eingenstates can be calculated straightforwardly from which the $z-$component of the Berry curvature reads
\begin{equation}
b^z({\bf p})=\pm \frac{1}{2}\frac{p_z\mp q/2}{|E({\bf p})|^3},
\end{equation}
with $E({\bf p})=\pm\sqrt{p^2+q^2/4\pm |{\bf p}\cdot{\bf q}|}$ \cite{AHEiQCD} and we have chosen, for simplicity, ${\bf q}=(0,0,q)$. 

\begin{figure}[htbp]
\includegraphics[scale=0.35]{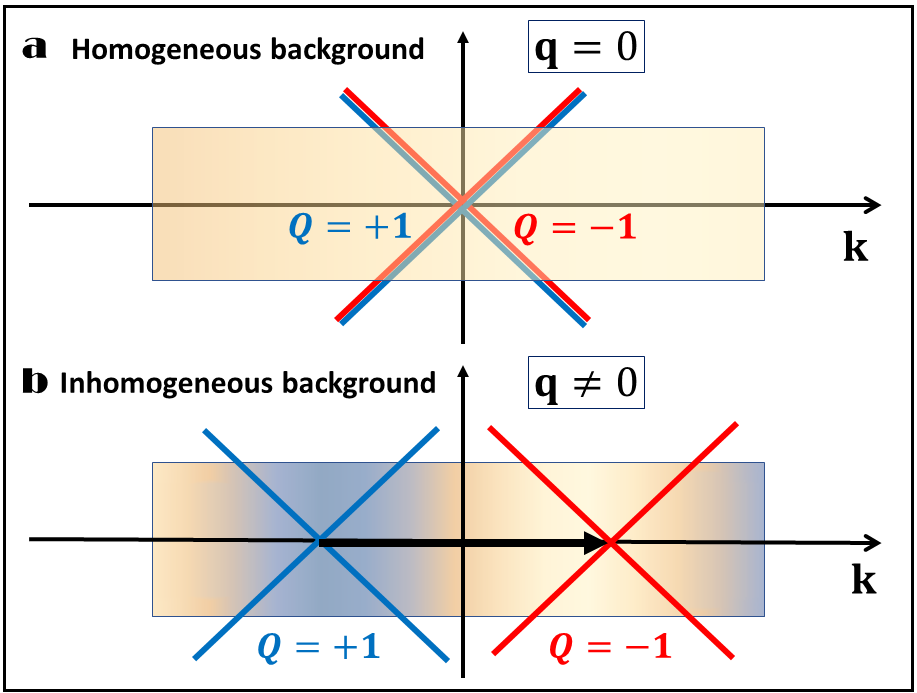}
\caption{{\bf a} $-$ For a homogeneous background, ${\bf q}=0$, the Weyl cones associated to the two topological charges, $Q=+1$ (right-handed screw axis) and $Q=-1$ (left-handed screw axis), are located at the ${\bf k}_{H,H^\prime}$ high-symmetry points in the Brillouin zone. Here, reciprocity forbids any anti-symmetric anomalous or planar Hall responses, as imposed by time-reversal symmetry. {\bf b} $-$ For an inhomogeneous background,  ${\bf q}\neq 0$, however, the two Weyl cones become shifted in momentum, ${\bf k}_\pm={\bf k}_{H}\pm\frac{{\bf q}}{2}$. Nonreciprocity thus allows for nonzero, anti-symmetric anomalous and planar Hall effects.}
\label{FigureInhomogeneousBackground}
\end{figure}

The above Berry curvature corresponds to magnetic monopoles shifted from the origin towards ${\bf p}=(0,0,\pm q/2)$, with opposite signs for the two chiral screw axis, see Fig. \ref{FigureInhomogeneousBackground}. This simple and general result demonstrates that spatial dispersion does indeed produce Weyl nodes that become shifted in momentum, breaking time-reversal symmetry, and thus allowing for both a ${\bf q}-$dependent anomalous Hall effect, as well as a ${\bf q}-$dependent chiral anomaly, which has already been acknowledged to be responsible for the planar Hall effect \cite{PHEWeylSemiMetals}. 

\subsection{Spatially dispersive, gyrotropic tellurium}

In order to check whether or not the above general argument applies to tellurium we must do a proper calculation, taking into account the full $4-$fold valence band structure calculated in ${\bf k}\cdot{\bf p}-$perturbation theory. Differently from the previous argument, we shall remain in the laboratory reference frame, in which both the Berry curvature and hedgehog texture remain are centered at the high-symmetry points, and introduce the spatial dispersion by extending the traditional, semiclassical approach of Boltzmann transport equations to the case of nonreciprocal transport. 

The current density, ${\bf j}({\bf r})$, the carrier distribution, $f({\bf k},{\bf r})$, and the velocity, $\dot{\bf r}$, satisfy ${\bf j}({\bf r})=-(e/V)\sum_{\bf k}f({\bf k},{\bf r})\dot{\bf r}$. Boltzmann's  equation in the steady-state, relaxation time, $\tau_{\bf k}$, approximation for $f({\bf k},{\bf r})$ reads \cite{PHEWeylSemiMetals,Johansson2019,XiaoRMP2010,Kim2014}
\begin{equation}
    \left(\dot{\bf r}\nabla_{\bf r}+\dot{\bf k}\nabla_{\bf k}\right)f({\bf k},{\bf r})=-
    (\tau_{\bf k}^{-1}\pm\tau_\phi^{-1})[f({\bf k},{\bf r})-f_{eq}({\bf k},{\bf r})],
    \label{BoltzmannEquation}
\end{equation}
where $f_{eq}({\bf k},{\bf r})$ is the equilibrium distribution and $+\tau_\phi^{-1}$ ($-\tau_\phi^{-1}$) represents the contribution from weak-localization (antilocalization) due to multiple impurity collisions \cite{Kim2014}. The equations of motion in the presence of Berry curvature are
\begin{equation}
    \dot{\bf r}=\frac{1}{\hbar}\nabla_{\bf k}\epsilon_{H_4^+}({\bf k})+
    \hbar\dot{\bf k}\times{\bf \Omega}_{\bf k},
    \quad
    \dot{\bf k}=\frac{e}{\hbar}{\bf E}+\frac{e}{\hbar c}\dot{\bf r}\times{\bf B}.  
    \label{EquationsofMotion}
\end{equation}

For a homogeneous system and as discussed in detail in the previous section, eq. (\ref{BoltzmannEquation}) can be expanded in powers of ${\bf E}$ and ${\bf B}$ to provide us with
\begin{widetext}
\begin{equation}
    {\bf j}=-\frac{e}{V}\sum_{\bf k}\eta_{\bf k}^{-1}
    \left(\tilde{\bf v}_{\bf k}+
    \frac{e}{\hbar}{\bf E}\times{\bf \Omega}_{\bf k}+
    \frac{e}{\hbar}({\bf \Omega}_{\bf k}\cdot\tilde{\bf v}_{\bf k}){\bf B}\right)
    \times
    \left\{\tilde{f}_{eq}({\bf k})+
    \frac{\partial \tilde{f}_{eq}({\bf k})}{\partial\tilde{\cal E}({\bf k})}
    \left[1+\frac{e(\tilde{\bf \Lambda}_{\bf k}\times{\bf B})\cdot\nabla_{\bf k}}{\hbar}\right]
    (e\tilde{\bf \Lambda}_{\bf k}\cdot{\bf E})\right\},
\end{equation}
\end{widetext}
where $\eta_{\bf k}=1+e{\bf \Omega}\cdot{\bf B}/\hbar$ is the curvature-modified density of states accounting for Fermi surface volume changes \cite{XiaoRMP2010}, and $\tilde{f}_{eq}({\bf k})$ is the Fermi-Dirac equilibrium distribution function for $\tilde{\cal E}({\bf k})=\epsilon_{H_4^+}({\bf k})-{\bf m}_{orb}({\bf k})\cdot{\bf B}$. 

\subsubsection{Chiral velocity shift - Berry curvature}

The induced Berry curvature, ${\bf \Omega}_{\bf k}$, calculated in eq. (\ref{3DBerryCurvature}) will provide a nontrivial contribution to the mean free path
\begin{equation}
\tilde{\bf \Lambda}_{\bf k}=\frac{\eta^{-1}_{\bf k}[\tilde{\bf v}_{\bf k}+e({\bf \Omega}_{\bf k}\cdot\tilde{\bf v}_{\bf k}){\bf B}/\hbar]}{(\tau_{\bf k}^{-1}\pm\tau_\phi^{-1})}
\label{MeanFreePath}
\end{equation}
which is given in terms of the relaxation time $\tau_{\bf k}$ and the chiral velocity shift $e({\bf \Omega}_{\bf k}\cdot\tilde{\bf v}_{\bf k}){\bf B}/\hbar$. This is the mechanism behind several chiral anomaly related phenomena, such as the chiral-magnetic and planar Hall effects \cite{PHEWeylSemiMetals}.

\subsubsection{Group velocity shift - hedgehog texture}

The $3D$ hedgehog orbital texture calculated in eq. (\ref{3DOrbitalMoment}) will provide a contribution to the group velocity shift
\begin{equation}
    \tilde{\bf v}_{\bf k}=\frac{1}{\hbar}\nabla_{\bf k}
    \left(\epsilon_{H_4^+}({\bf k})-{\bf m}_{orb}({\bf k})\cdot{\bf B}\right)=
    {\bf v}_{\bf k}+\delta{\bf v}_{\bf k},
    \label{GroupVelocityShift}
\end{equation}
for the charge carriers at the valence band. The group velocity shift $\delta{\bf v}_{\bf k}=-\nabla_{\bf k}({\bf m}_{orb}\cdot{\bf B})$ has been previously considered in the description of the magnetochiral anisotropy in $p-$type tellurium \cite{Rikken2019}. For ${\bf B}\perp{\bf j}\parallel\hat{\bf z}$ there is an energy splitting arising from the breakdown of the degeneracy between valence band states with ${\bf k}$ and $-{\bf k}$ \cite{Doi1970,Nakao1971}, that needs to be enantioselective in order to be symmetry allowed. For this reason, in ref. \cite{Rikken2019} it was proposed, heuristically, that such energy splitting should be $\Delta{\epsilon}({\bf k})=\chi^{D/L}{\bf k}\cdot{\bf B}$, with $\chi^D=-\chi^L$. Our eq. (\ref{GroupVelocityShift}), together with eqs. (\ref{z-orbital-moment}) and (\ref{3DOrbitalMoment}), thus places such heuristic argument on solid theoretical grounds since the coupling ${\bf m}_{orb}\cdot{\bf B}$ is inherently enantioselective. 

\subsection{Reciprocal conventional Hall effect}

The Lorentz force generates a reciprocal Hall current corresponding to the totally anti-symmetric, linear in $B$ term, $h(\omega)\epsilon_{ijk}B_k$ from eq. (\ref{ReciprocalOnsager}) for the Onsager's reciprocal relations 
\begin{equation}
     {\bf j}=-\frac{e^3}{V\hbar}\sum_{\bf k}
     \frac{\eta_{\bf k}^{-3}\tilde{\bf v}_{\bf k}}{(\tau_{\bf k}^{-1}\pm\tau_\phi^{-1})^2}
     \frac{\partial \tilde{f}_{eq}({\bf k})}{\partial\tilde{\cal E}({\bf k})}
    (\tilde{\bf v}_{\bf k}\times{\bf B})\cdot\nabla_{\bf k}(\tilde{\bf v}_{\bf k}\cdot{\bf E}),
\end{equation}
and a Hall conductivity tensor 
\begin{equation}
\sigma_{zx}^{h}\sim i\epsilon_{zxy} B_y\sim B\cos\theta.
\label{ReciprocalHall}
\end{equation}
Because of time reversal symmetry, however, a reciprocal anomalous Hall current, 
\begin{equation}
    {\bf j}=-\frac{e}{V}\sum_{\bf k}\eta_{\bf k}^{-1}({\bf E}\times{\bf \Omega}_{\bf k})\tilde{f}_{eq}({\bf k})=0,
    \label{ReciprocalAHE}
\end{equation} 
vanishes. The reason for the vanishing of eq. (\ref{ReciprocalAHE}) is very simple. Time reversal symmetry ensures that the Berry curvature is odd under ${\bf k}\rightarrow -{\bf k}$
\begin{equation}
    {\bf \Omega}_{-{\bf k}}=-{\bf \Omega}_{{\bf k}}.
\end{equation}
Since the sum in eq. (\ref{ReciprocalAHE}) is for the entire Brillouin zone, including positive as well as negative values for the Berry curvature, then eq. (\ref{ReciprocalAHE}) vanishes identically. The same holds for the reciprocal PHE \cite{Johansson2019}, or simply
\begin{equation}
{\bf j}=-\frac{e}{V}\sum_{\bf k}\eta_{\bf k}^{-2}\tilde{\bf v}_{\bf k}
    ({\bf \Omega}_{\bf k}\cdot\tilde{\bf v}_{\bf k})
    \partial \tilde{f}_{eq}({\bf k})/\partial\tilde{\cal E}({\bf k})
    ({\bf B}\cdot{\bf E})=0.  
    \label{ReciprocalPHE}
\end{equation}
Once again, not only ${\bf \Omega}_{\bf k}$ is odd under time reversal, but so is the velocity, $\tilde{\bf v}_{-{\bf k}}=-\tilde{\bf v}_{{\bf k}}$. As a result, the overall parity of the term $\tilde{\bf v}_{\bf k}({\bf \Omega}_{\bf k}\cdot\tilde{\bf v}_{\bf k})$ is odd and for that reason eq. (\ref{ReciprocalPHE}) also vanishes identically. Time reversal symmetry forbids both reciprocal AHE and reciprocal PHE.

It is important to emphasize, at this point, that the absence of a reciprocal, ${\bf q}=0$, AHE occurs only at the linear order in the electric field, ${\bf E}$, see eq. (\ref{ReciprocalAHE}). However, reciprocal, ${\bf q}=0$, anomalous Hall-like currents do indeed arise as quantum nonlinear effects, ${\bf E}^2$, induced by Berry curvature dipole tensor, $D_{ab}=\int_{\bf k} f_{eq}({\bf k})\partial_a{\bf \Omega}_b({\bf k})\neq 0$, in noncentrosymmetric materials, even in the presence of time reversal symmetry \cite{Sodemann2015}. This remarkable reciprocal, nonlinear Hall effect has a purely quantum origin and produces an anomalous velocity shift when the system is in a current carrying state \cite{Sodemann2015}. 

\subsection{Nonreciprocal Response}

In what follows, we shall demonstrate that both the anomalous and planar Hall effects arise in the linear electric field regime, ${\bf E}$, and nonreciprocal response, ${\bf q}\neq 0$. For gyrotropic media such as tellurium, two novel, nonreciprocal Hall responses are possible. The electrical acceleration $\delta\dot{\bf k}$ imposed to the carriers by an AC current follows the same  $e^{i({\bf q}\cdot{\bf r}-\omega t)}$ dependence \cite{DresselGruener}, so
\begin{equation}
    \delta\dot{\bf k}=\frac{e{\bf E}/\hbar}
    {1-i\omega\tau_{\bf k}+i(\tilde{\bf v}_{\bf k}\cdot{\bf q})\tau_{\bf k}}.
\end{equation}

\subsubsection{AHE - group velocity shift - hedgehog texture}

This leads to an anomalous Hall contribution at zero-th order in the expansion of the distribution function \cite{Zhong2015} originated from the group velocity shift $\delta{\bf v}_{\bf k}=-\nabla_{\bf k}({\bf m}_{orb}\cdot{\bf B})/\hbar$ 
\begin{equation}
    {\bf j}=-\frac{e^2}{V\hbar}\sum_{\bf k}\eta_{\bf k}^{-1}
    \left[\frac{\delta{\bf v}_{\bf k}\cdot{\bf q}}{\omega}\right]
    \left({\bf E}\times{\bf \Omega}_{\bf k}\right)\tilde{f}_{eq}({\bf k}),
    \label{NovelHall}
\end{equation}
whose linear in $B$ dependence at small fields is stabilized by the wave-vector components ${q}_m=\omega {n}_m/c$ for the spatially dispersive, anisotropic media with refraction index ${\bf n}$. Differently from the vanishing reciprocal AHE in eq. (\ref{ReciprocalAHE}), now the ${\bf q}-$dependent term inside the brackets in eq. (\ref{NovelHall}) brings a second odd quantity, the group velocity shift, $\delta{\bf v}_{-{\bf k}}=-\delta{\bf v}_{\bf k}$, to the sum in the Brillouin zone, ensuring that the nonreciprocal anomalous Hall response does not vanish in the presence of the Berry curvature, ${\bf \Omega}_{\bf k}$, which is also odd under ${\bf k}\rightarrow -{\bf k}$. 

We shall neglect corrections from the density of states, $\eta_{\bf k}$, and drop all integral over occupied states of $({\bf \Omega}_{\bf k}\cdot\tilde{\bf v}_{\bf k}){\bf B}$, because these are known to vanish (absense of an equilibrium chiral magnetic effects) even in the presence of monopole singularities \cite{Zhong2015}. For gyrotropic, spatially dispersive media Eq. (\ref{NovelHall}) can be rewritten in terms of the constitutive equations 
${j}_i=\gamma_{ik\ell}\partial_\ell {E}_k=
    -i\epsilon_{ik\ell}g_{\ell m}({\bf B}){E}_k{q}_m$ \cite{LandauLifshitz},
Because the tensor $\gamma_{ik\ell}$ is antisymmetric, $\gamma_{ik\ell}=-\gamma_{ki\ell}$, one often makes use of the dual pseudotensor, $\epsilon_{ikm}g_{m\ell}=\omega\gamma_{ik\ell}/c$, 
where 
\begin{equation}
    g_{m\ell}({\bf B})=\frac{i e^2}{V\hbar\omega}\sum_{\bf k}\eta_{\bf k}^{-1}
    f_{eq}({\bf k})\delta{v}^\ell_{\bf k}({\bf B}){\Omega}^m_{\bf k},
    \label{Rank-two-g}
\end{equation}
is only nonzero for systems lacking an inversion center \cite{Zhong2015}. 

\begin{figure*}
\includegraphics[scale=0.5]{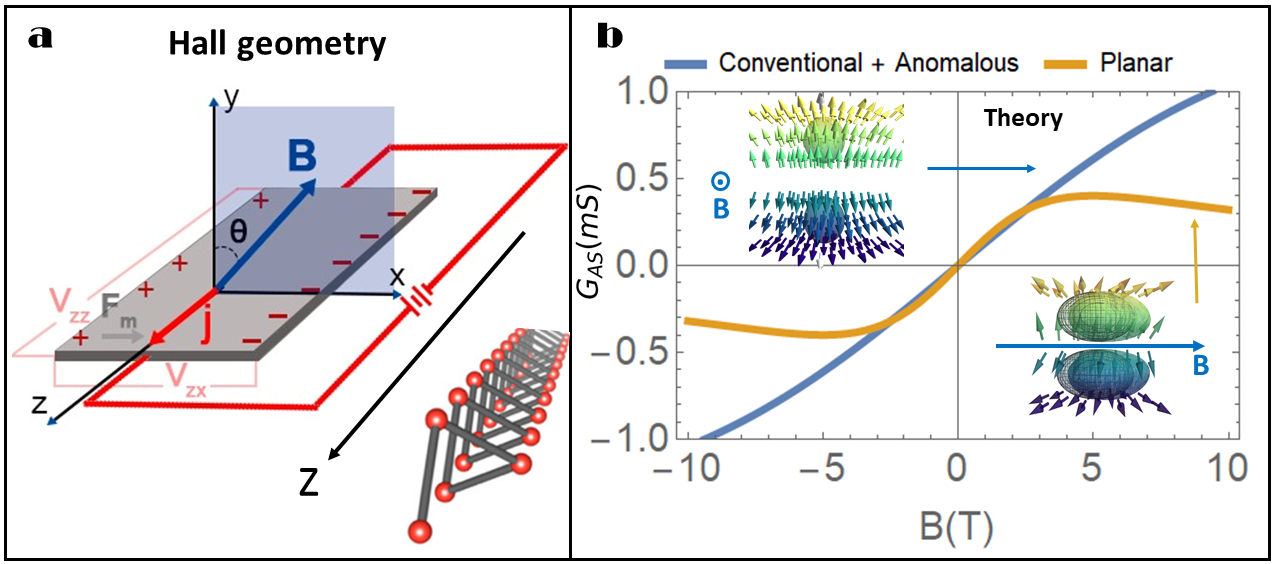}
\caption{{\bf a} $-$ Hall geometry ${\bf j}\parallel\hat{\bf z}\perp{\bf B}\not\parallel{\bf E}\parallel\hat{\bf x}$; {\bf b} $-$ theoretical curves for the conventional $+$ anomalous and planar Hall responses. Inset: effects of $B$ at $\theta=0$, top left, and $\theta=\pi/2$, bottom right, to the orbital texture.}
\label{FigureGeometryTheo}
\end{figure*}

For ${\bf j}\parallel\hat{\bf z}\perp{\bf B}\not\parallel{\bf E}\parallel\hat{\bf x}$ as in Fig. \ref{FigureGeometryTheo}$-$a the novel, nonreciprocal AHE requires that both the induced hedgehog orbital texture, $m_{orb}^y({\bf k})\neq 0$, as well as the Berry curvature, ${\Omega}^y_{\bf k}\neq 0$, have nonzero $y$ components, which can only be the case due to the presence of a ${\bf k}-$dependent spin-orbit interaction. In this case, for small $B$ the novel, nonreciprocal AHE reads
\begin{equation}
\tilde{\sigma}^{ah}_{zx}\sim B\cos\theta,
\end{equation}
and adds up to the reciprocal contribution from the conventional (Lorentz force) Hall effect given in eq. (\ref{ReciprocalHall}).

\subsubsection{PHE - chiral velocity shift - Berry curvature}

Since we are looking for the possibility of a linear in $B$ planar Hall effect, from start we have intentionally chosen ${\bf j}\perp{\bf B}$ (or $\varphi=\pi/2$) to remove any possible symmetric $B^2\sin(2\varphi)$ PHE contribution from the Hall response \cite{Zhang2020}. In this case, a nonreciprocal, antisymmetric PHE can also be stabilized in spatially dispersive media by the chiral velocity shift $\delta{\bf v}_{\bf k}=e({\bf \Omega}_{\bf k}\cdot\tilde{\bf v}_{\bf k}){\bf B}/\hbar$
\begin{equation}
    {\bf j}=-\frac{e^3}{V\hbar}\sum_{\bf k}
    \frac{\eta_{\bf k}^{-2}\tilde{\bf v}_{\bf k}}{\tau_{\bf k}^{-1}\pm\tau_\phi^{-1}}
    \left[\frac{\tilde{\bf v}_{\bf k}\cdot{\bf q}}{\omega}\right]
    \frac{\partial \tilde{f}_{eq}({\bf k})}{\partial\tilde{\cal E}({\bf k})}
    ({\bf \Omega}_{\bf k}\cdot\tilde{\bf v}_{\bf k})({\bf B}\cdot{\bf E}),
    \label{NonRecPHE}
\end{equation}
and provides us with $j_{i}=g_{ik\ell m}B_\ell q_m E_k=D_{im}\delta_{\ell k} B_\ell q_m E_k$, where the even-rank chiral diffusion pseudotensor \cite{OnsagerNonReciprocal}
\begin{equation}
    D_{i m}=\frac{e^3}{V\hbar\omega}\sum_{\bf k} 
    \frac{\eta_{\bf k}^{-2}}{\tau_{\bf k}^{-1}\pm\tau_\phi^{-1}}
    \frac{\partial \tilde{f}_{eq}({\bf k})}{\partial\tilde{\cal E}({\bf k})}
    v^i_{\bf k}({\bf \Omega}_{\bf k}\cdot{\bf v}_{\bf k}) v^m_{\bf k}.
    \label{Rank-four-g}
\end{equation}
Once again the ${\bf q}-$dependent term in eq. (\ref{NonRecPHE}) ensures the sum over ${\bf k}$ in the Brillouin zone not to vanish, because even though ${\bf v}_{\bf k}$ and ${\bf \Omega}_{\bf k}$ individually are odd, the four-product $v^i_{\bf k}({\bf \Omega}_{\bf k}\cdot{\bf v}_{\bf k}) v^m_{\bf k}$ is even under time reversal. In this case, for ${\bf j}\parallel\hat{\bf z}\perp{\bf B}\not\parallel{\bf E}\parallel\hat{\bf x}$ as in Fig. \ref{FigureGeometryTheo}$-$a a nonreciprocal PHE requires that ${\Omega}^{(x,y)}_{\bf k}\neq 0$ and is therefore our second evidence of a hedgehog texture with 
\begin{equation}
\tilde{\sigma}^{ph}_{zx}\sim B\sin\theta.
\end{equation}
Most importantly, the nonreciprocal PHE has a different phase than both the reciprocal Hall and nonreciprocal AHE. The PHE is distinguishable from the other two and will leave unambiguous signatures in transport, to be discussed below.

\begin{figure*}
\includegraphics[scale=0.5]{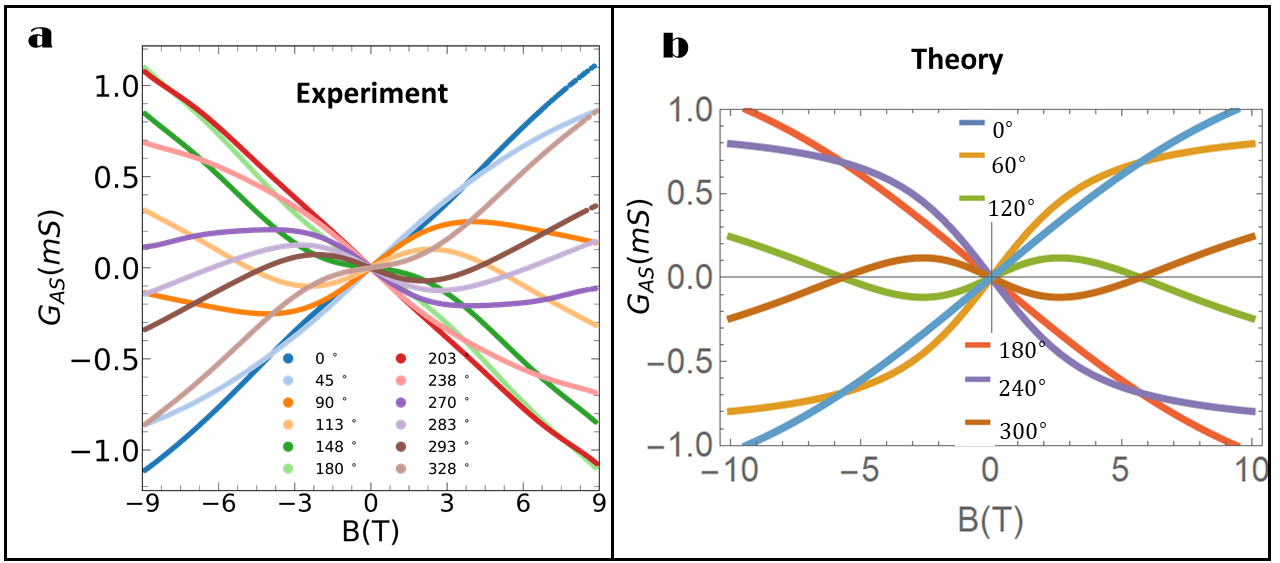}
\caption{{\bf a} $-$ Experimental and {\bf b} $-$ Theoretical curves for the antysimmetric AC magnetoconductance as a function of the magnetic field, $B$, for various angles $0\leq\theta\leq 2\pi$.}
\label{FigureTheoryExperiment}
\end{figure*}

\section{Comparison with Experiments}

In order to test our predictions, we performed magnetoconductance (MC) measurements at room temperature on $p-$type Sn-doped tellurium and extracted the antisymmetric component $G_{AS}$ from the data. The sample has been grown by slow cooling from a fine tellurium powder with 0.04\% added antimony. We have measured the response for different orientations of the external field, which has been rotated within the $xy$ plane. Contacts with platinum wires were glued with silver ink and prepared in a classical 4-point set-up. As a measurement device, we use a commercial PPMS Dynacool with a maximum field of 9 T and a minimum temperature of 1.8 K from Quantum Design. The samples have been installed on the horizontal rotator stage. The current direction is always along the $z$-axis direction and the Quantum Design ETO-option was used for AC-transport data acquisition. Naturally, the antisymmetric MC-component is proportional to the Hall response as in a Hall geometry shown in Fig. \ref{FigureGeometryTheo}$-$a. 

\begin{figure*}
\includegraphics[scale=0.5]{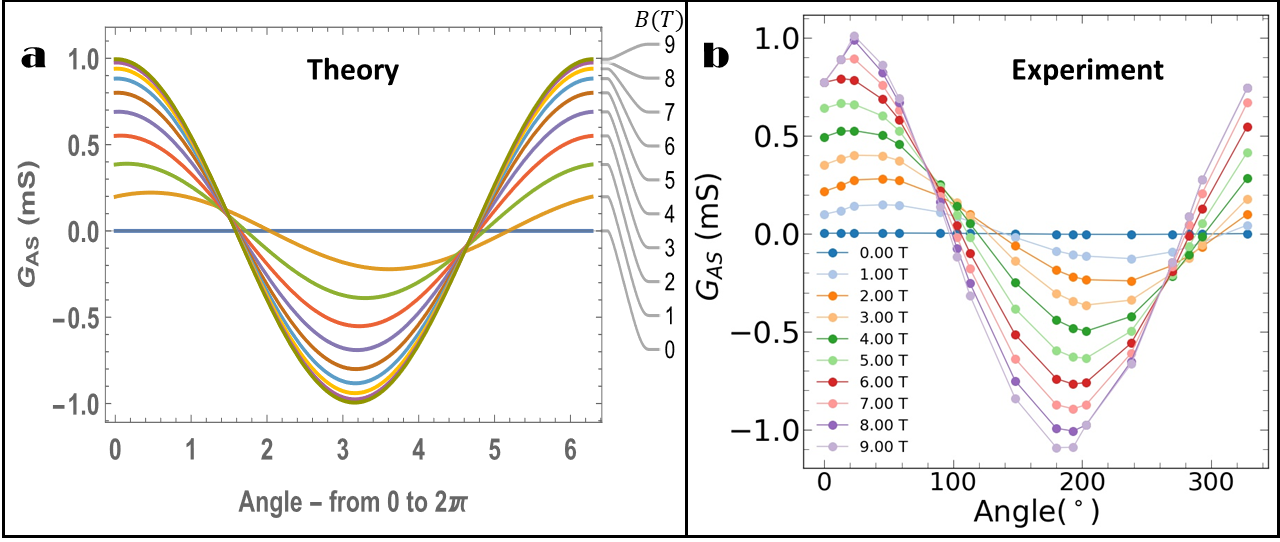}
\caption{{\bf a} $-$ Theoretical and {\bf b} $-$ Experimental curves for the antysimmetric AC magnetoconductance as a function of the angle, $\theta$, for various intensities of the applied magnetic field, $0\leq B\leq 9$T.}
\label{FigureTheoryExperimentTheta}
\end{figure*}

In \ref{FigureGeometryTheo}$-$b, we show our predictions for the conventional, anomalous and planar Hall responses. For $\theta=0$, the almost linear behavior originates from the conventional (reciprocal) and anomalous (nonreciprocal) Hall signals (both indistinguishable) in the regime $\omega_c\tau\ll 1$, where $\omega_c=e B/m^*c$ is the cyclotron frequency and $\tau^{-1}=\tau_0^{-1}+\tau_\phi^{-1}$ is a lower bound for the multiple-impurity collisions, weak-localization relaxation time. For $\theta=\pi/2$, however, the non-monotonic behavior corresponds to the planar (nonreciprocal) Hall effect in the regime $\omega_c\tilde{\tau}\sim 1$, where $\tilde{\tau}\approx\tau_0$ is an upper bound for the single-impurity, $\tau_\phi^{-1}\rightarrow 0$, relaxation time \cite{Kim2014}. Clearly the alignment of the orbital magnetic moments, ${\bf m}_{orb}\parallel\hat{\bf x}$, promoted by an applied field ${\bf B}\parallel\hat{\bf x}$, results in the phase-coherence among the scattering states with ${\bf k},{\bf k}^\prime$ in the relevant $zx$ Hall geometry of Fig. \ref{FigureGeometryTheo}$-$a), suppressing multiple-impurity scattering and enlarging the relaxation time. 
Obviously, the Hall response at planar field orientation should be zero, without any non-trivial effects. This  becomes even clearer, when taking the derivative of $G_{AS}$ for different external field strength and plot it vs. the applied field angle, as shown in Fig. \ref{FigureGeometryTheo} b). 
As the magnetic field rotates from $0\leq\theta\leq 2\pi$, the antisymmetric contribution to the $zx$ Hall conductivity, $\sigma_{zx}^{AS}$, shown in Figs. \ref{FigureTheoryExperiment} and \ref{FigureTheoryExperimentTheta} can be remarkably well described by
\begin{equation}
    \sigma_{zx}^{AS}=(\sigma_{zx}^{h}+\tilde{\sigma}_{zx}^{ah})
    \frac{\omega_c\tau\,\cos\theta}{1+(\omega_c\tau)^2}+
    \tilde{\sigma}_{zx}^{ph}
    \frac{\omega_c\tilde{\tau}\,\sin\theta}{1+(\omega_c\tilde{\tau})^2},
    \label{TotalHallResponse}
\end{equation}
where the ratios $\alpha=(\sigma_{zx}^{h}+\tilde{\sigma}_{zx}^{ah})/\tilde{\sigma}_{zx}^{ph}$ and $\beta=\tilde{\tau}/\tau$ are the same for all curves. For $\tan{\theta}>0$ we find $\sigma_{zx}^{AS}=0$ only at $\omega_c=0$, that is, at zero field, $B=0$. For $\tan{\theta}<0$, however, besides $\omega_c=0$, the conductivity $\sigma_{zx}^{AS}=0$ also at
$\omega_c\approx\pm\sqrt{\beta|\tan{\theta}|/\alpha - 1}/\beta\tau$ for $\omega_c\tau\ll 1$, see Fig. \ref{FigureTheoryExperimentTheta}.

It is remarkable that the quantum corrections to conductivity, encoded in the weak-localization relaxation time $\tau_\phi$, $\beta\gg 1$, are accessible to our AC measurements even at $T=300$K. By looking at the graph, we see that for high fields the phase is located around zero degrees, proving that within this field range the perpendicular Hall signal dominates. Unfortunately the contributions from the conventional Hall and nonreciprocal anomalous Hall effects are not distinguishable within the sample set we used. In the low field, on the other hand, the phase is shifted by about $50^{\circ}$, proving the existence of the planar component in the data. Here the existence of an anti-symmetric component for the planar Hall current is evident. The field dependence of the phase shift is not due to any misalignment of the contacts, that would produce an overall constant phase shift irrespective of the field intensity. Finally, the two novel, nonreciprocal Hall responses reported in this work are linear in $B$ for small fields, since for carrier concentrations of the order $n\sim 10^{16}/cm^{3}$ the Fermi energy calculated from the free electron model gives us $E_F\sim 0.1$meV, which corresponds to a Zeeman field of $~10$T. 

\section{Conclusions} 

In conclusion, our work unveils the true origin of the Weyl physics observed in gyrotropic, $p-$type tellurium by demonstrating how a $3D$ hedgehog chiral magnetic texture is induced at the topologically trivial top valence bands by their $k-$SOI to two Weyl node containing  bands. Therewith, our findings not only clarify the origin of previous experimental observations of Weyl signatures in tellurium \cite{Zhang2020}, even though Weyls nodes are not directly accessible \cite{Hirayama2015}, but also bring into light two novel, nonreciprocal and antisymmetric Hall responses allowed in spatially dispersive media with natural optical activity, and accessed via AC measurements. Finally, our work opens up new venues for scientific and technological research by including nonreciprocity into the holography dictionary, as a candidate route for generating axial components to dual gauge fields \cite{Gauge-Gravity-Holography}, lays the foundations for tellurium based, nonreciprocal topological FETs \cite{TelluriumFET}, and also introduces alternative controlling mechanisms for AC electron transfer in enantioselective molecules \cite{EnantioSelectivityChemistry}.

\section{Acknowledgements} 

The authors acknowledge the financial support from CAPES and CNPq. EBS acknowledges CNPq and several grants from FAPERJ, including Professor Emeritus fellowship. JF acknowledges FAPERJ for his PDR 10.

\end{document}